\documentclass[11pt]{article}
\pdfoutput=1
\usepackage{dcolumn}
\usepackage{bm}

\usepackage{graphicx}
\usepackage{amssymb,amsmath}
\usepackage{multirow}
\usepackage{cite,color,url}
\usepackage[colorlinks=true
,urlcolor=blue
,anchorcolor=blue
,citecolor=blue
,filecolor=blue
,linkcolor=blue
,menucolor=blue
,linktocpage=true
,pdfproducer=medialab
,pdfa=true
]{hyperref}

\usepackage{slashed}
\usepackage{epsfig,psfrag,rotating,soul}
\usepackage{rotfloat}

\evensidemargin \oddsidemargin
\allowdisplaybreaks

\let\OLDthebibliography\thebibliography
\renewcommand\thebibliography[1]{
  \OLDthebibliography{#1}
  \setlength{\parskip}{0pt}
  \setlength{\itemsep}{0pt plus 0.3ex}
}

\def\pvslash{\vec{p}\hspace{-10pt}\not{}\hspace{2pt}}
\def\pvsslash{\vec{p}\hspace{-5pt}\not{}\hspace{2pt}}
\begin{document}
\thispagestyle{empty}

\def\thefootnote{\fnsymbol{footnote}}

\begin{flushright}
BONN-TH-2018-06\\
IFT-UAM/CSIC-18-037\\
\end{flushright}

\vspace*{1cm}

\begin{center}

\begin{Large}
\textbf{\textsc{Potential discovery of staus \\[.25em] through heavy Higgs boson decays at the LHC}}
\end{Large}

\vspace{1cm}

{\sc
Ernesto~Arganda$^{1, 2}$%
\footnote{{\tt \href{mailto:ernesto.arganda@fisica.unlp.edu.ar}{ernesto.arganda@fisica.unlp.edu.ar}}}%
, Victor~Mart\'{	\i}n-Lozano$^{3}$%
\footnote{{\tt \href{mailto:lozano@th.physik.uni-bonn.de}{lozano@th.physik.uni-bonn.de}}}, Anibal D.~Medina$^{1}$%
\footnote{{\tt \href{mailto:anibal.medina@fisica.unlp.edu.ar}{anibal.medina@fisica.unlp.edu.ar}}}%
and Nicolas~I.~Mileo$^{1}$%
\footnote{{\tt \href{mailto:mileo@fisica.unlp.edu.ar}{mileo@fisica.unlp.edu.ar}}}%

}

\vspace*{.7cm}

{\sl
$^1$IFLP, CONICET - Dpto. de F\'{\i}sica, Universidad Nacional de La Plata, \\ 
C.C. 67, 1900 La Plata, Argentina

\vspace*{0.1cm}

$^2$Instituto de F\'{\i}sica Te\'orica UAM/CSIC, \\ Calle Nicol\'as
Cabrera 13-15, Cantoblanco E-28049 Madrid, Spain

\vspace*{0.1cm}

$^3$Bethe Center for Theoretical Physics \& Physikalisches Institut der Universit\"at Bonn,\\ Nu{\ss}allee 12, 53115, Bonn, Germany
}

\end{center}

\vspace{0.1cm}

\begin{abstract}
\noindent
In this work we present a new search strategy for the discovery of staus at the LHC in the context of the minimal supersymmetric standard model. The search  profits from the large s-channel $b$-quark annihilation production of the heavy CP-even and CP-odd Higgs bosons ($H/A$) which can be attained in regions of $\tan\beta\gg 1$ that avoid the stringent $H/A \to \tau^+ \tau^-$ searches via decays into stau pairs.  We also focus on regions where the  staus branching ratios are dominated by the decays into a tau lepton and the lightest neutralino. Thus the experimental signature  consists of final states made up of a tau-lepton pair plus large missing transverse energy. We take advantage of the large stau-pair production cross sections via heavy Higgs boson decays, which are between one or two orders of magnitude larger than the usual electroweak production cross sections for staus. A set of basic cuts allow us to obtain significances of the signal over the SM backgrounds at the discovery level (5 standard deviations) in the next LHC run with a center-of-mass energy of 14 TeV and a total integrated luminosity of only 100 fb$^{-1}$.

\end{abstract}

\def\thefootnote{\arabic{footnote}}
\setcounter{page}{0}
\setcounter{footnote}{0}

\newpage

\section{Introduction}
\label{intro}

Supersymmetry (SUSY) is a well studied theory beyond the standard model (SM) of particle physics (for a review, see, e.g.,~\cite{Martin:1997ns}).  In its minimal version, the minimal supersymmetric standard model (MSSM) (for reviews, see, e.g.,~\cite{Nilles:1983ge,Haber:1984rc}) elegantly solves the gauge hierarchy problem via the introduction of additional particles (superpartners) with opposite statistics to those of the SM and can acomodate a SM-like Higgs with a 125 GeV mass, as measured at the Large Hadron Collider (LHC)~\cite{Aad:2012tfa,Chatrchyan:2012xdj,Aad:2015zhl}. Great effort is being put forward at the LHC in the pursuit of signals of SUSY and the lack of any quantitative deviation from the SM expectations has led the particle physics community to consider {\it natural} SUSY spectrums in which first and second generation squarks and sleptons are somewhat decoupled from the low-energy effective theory~\cite{Dimopoulos:1995mi,Pomarol:1995xc,Cohen:1996vb}. Furthermore, the MSSM Higgs spectrum consists of a type-II two Higgs doublet model, which in the CP-conserving case can be decomposed into two CP-even Higgs bosons $h$ and $H$, one CP-odd Higgs boson $A$ and a charged Higgs boson pair $H^{\pm}$. The lightest CP-even Higgs is usually identified with the 125 GeV scalar resonance discovered at the LHC~\cite{Heinemeyer:2011aa}, whereas there are current searches for $H$ and $A$ in the di-tau channel~\cite{Chatrchyan:2011nx,Aad:2011rv,Chatrchyan:2012vp,Aad:2012cfr,Khachatryan:2014wca,Aaboud:2016cre,Aaboud:2017sjh,Sirunyan:2018zut} which tend to provide strong constraints in the [$m_A$, $\tan\beta$] plane, in particular in the large $\tan\beta$ and small $m_A$ region.  The reason behind this is that the $H$ coupling to down-type fermions grows with $\tan\beta$. This region of large $\tan\beta$ is very interesting since  it leads to Yukawa coupling unification $y_t\approx y_b\approx y_{\tau}$ and furthermore naturally provides SM-like properties for $h$. It was shown in~\cite{Medina:2017bke} that by considering SUSY decays of $H$ into third generation squarks and sleptons, partial portions of the large $\tan\beta$ regions could avoid the constraints from $H/A \to \tau^+ \tau^-$ and be consistent with all other experimental constraints. For third generation down-type sfermions, this was accomplished by an increment in  $A_d$, the chiral coupling via the trilinear soft-breaking terms in the potential. In this work we take advantage of these regions for the case of heavy CP-even Higgs decays into staus, for which the BR$(H\to\sum_{i,j=1,2}{\tilde{\tau}}^{*}_i\tilde{\tau}_j)$ can be as large as $\sim$ 0.25. Light staus in the sub-TeV region can be naturally obtained  for example in standard gaugino-mediated scenarios~\cite{Heisig:2017lik} and constraints on their masses from collider searches still allow them to be as low as 100 GeV. Interestingly, it turns out that in the newly allowed regions of large $\tan\beta$,  the production of stau pairs via heavy Higgs boson decays can be substantially larger than the usual considered electroweak (EW) pair production, which is currently the main search channel at the LHC~\cite{Aad:2014yka,Aad:2015eda,ATLAS:2016PUB,CMS:2017wox,CMS:2017rio}.

In this work, we consider the production of stau pairs as the decay products of the resonant heavy Higgs CP-even scalar $H$, with both scalar fermions decaying subsequently into a tau lepton and the lightest neutralino that is taken as the lightest supersymmetric particle (LSP). The final state is made up of two opposite-sign (OS) tau leptons and large missing transverse energy ($E_T^{\text{miss}}$) originated from the pair of LSP neutralinos. For the topology of these final states, a powerful discriminating variable commonly used is  $m_{T2}$, which depends on the momenta of the two visible particles and the $E_T^{\text{miss}}$ present in the event. We show that this variable $m_{T2}$ is extremely useful in order to discriminate signal from the main SM backgrounds and, together with other basic cuts, provides significances at the discovery level (5 standard deviation) for total integrated luminosities as small as 100 fb$^{-1}$ at the LHC. These  luminosity values  will be probed in the near future by this hadron collider~\cite{LHCprospects}.

The paper is organized as follows: The main theoretical features and phenomenological implications of the MSSM scenarios with large stau mixing are summarized in Section~\ref{theory}. Our alternative search strategy for staus at the LHC is presented in Section~\ref{analysis}, together with a general interpretation of the collider analysis within the large stau-mixing MSSM scenario. Finally, Section~\ref{conclusions} is devoted to a general discussion of the results and to present the main conclusions.

\section{The MSSM with large stau mixing}
\label{theory}

The mixed-chirality couplings of the heavy CP-even Higgs $H$ and the CP-odd Higgs $A$ to down-type sfermions for $\tan\beta\gg1$ take the form
\begin{eqnarray}
g_{A \tilde{d}_{L}\tilde{d}_{R}} &=&-\frac{1}{2}m_d\left[\mu +A_{d}\tan\beta\right],\qquad \;\;\;\;g_{H \tilde{d}_{L}\tilde{d}_{R}}=-\frac{1}{2}m_d\left[-\mu +A_{d}\tan\beta\right] \,.
\label{Eq:chiralcoupl}
\end{eqnarray}
These couplings grow in the same fashion with $\tan\beta$ as the couplings to down-type fermions and, at the same time, increase with $A_d$. Thus, if the mixing among the chiral sfermions states is close to maximal, it becomes possible to decrease the BR$(H/A\to \tau^+ \tau^-)$ by increasing specifically the BR$(H\to\sum_{i,j=1,2}{\tilde\tau}^{*}_i\tilde{\tau}_j)$.

In~\cite{Medina:2017bke}, a study on the possibility of enlarging the allowed regions in the [$m_A$, $\tan\beta$] plane through SUSY decays of the heavy Higgs bosons $H/A$ was performed. More precisely, the authors show that the severe constraints arising from searches for $H/A$ decaying into tau-lepton pairs can be avoided by considering new decays to light third-generation sfermions (see Figure 6 of~\cite{Medina:2017bke}). The details of the scanning procedure are fully discussed in~\cite{Medina:2017bke} and we urge the reader to review them. In this study, we focus on the orange points that appear as a strip on the left hand side of Figure 6 of~\cite{Medina:2017bke}, covering the range of $m_H\approx m_A$ between $[0.8, 1.2]$ TeV and $\tan\beta\in[25, 50]$, for which the existing di-tau constraints are evaded.

For the mentioned set of points, the cross sections for $H$ production via $b$-quark annihilation, $\sigma_{bbH}$, were found to be in the range of approximately $0.1$ pb (with some small variation depending on $m_H$) leading to $\sigma_{bbH}\times$BR$(H\to\sum_{i,j=1,2}{\tau}^{*}_i\tilde{\tau}_j)\lesssim 10^{-2}$ pb. In addition, we have analyzed the heavy CP-even Higgs decays and found that there are basically two situations for which the branching ratio into staus can be sizable: in one of them the dominant decay mode is $\tilde{\tau}^{*}_{1}\tilde{\tau}_{1}$ while in the other the decay mode into $\tilde{\tau}^{*}_{1}\tilde{\tau}_2+c.c$ dominates. In order to clarify this statement, let us briefly review the stau mass matrix and mixings:
%
\begin{eqnarray}
\left(\!\!
\begin{array}{ccc}
m_{\tilde{\tau}_{11}}  & m_{\tilde{\tau}_{12}}   \\
m_{\tilde{\tau}_{12}}  & m_{\tilde{\tau}_{22}}
\end{array}
\!\!\right)\!\!\!\!
&\equiv&
\!\!\!\!\left(\!\!
\begin{array}{ccc}
m^2_{L_{3}}+m_{\tau}^2+\left(-\frac{1}{2}+\frac{1}{3}\sin^2\theta_W\right)m_Z^2\cos2\beta  & m_{\tau}(A_{\tau}- \mu \tan\beta)  \\
 .    & m^2_{E_{3}}+m_{\tau}^2-\frac{1}{3}\sin^2\theta_W m_Z^2\cos2\beta 
\end{array}
\!\!\right),\nonumber\\
\end{eqnarray}
where we assumed $m^{*}_{\tilde{\tau}_{12}}=m_{\tilde{\tau}_{12}}$ and that the $A_{\tau}$ term is defined in such a way that $\Delta \mathcal{L}_\text{soft}=y_{\tau} A_{\tau} \tilde{\bar{e}}_{3}\tilde{L}_{3}H_{d}+c.c.$.  Diagonalizing the mass matrix, one obtains the mass eigenstates  $\tilde{\tau}_1$ and $\tilde{\tau}_2$  with masses given by
\begin{equation}
m_{\tilde{\tau}_2,\tilde{\tau}_1}^2=\frac{1}{2}(m_{\tilde{\tau}_{11}}+m_{\tilde{\tau}_{22}}\pm\Delta) \,,
\label{eq:mstau}
\end{equation}
with $\Delta \equiv \sqrt{(m_{\tilde{\tau}_{11}}-m_{\tilde{\tau}_{22}})^2+4 m^2_{\tilde{\tau}_{12}}}$. The  mixing angle between the flavour states is roughly given by  $\theta_\text{mix}\sim m_{\tilde{\tau}_{12}}/(m_{\tilde{\tau}_{11}}-m_{\tilde{\tau}_{22}})$. As can be seen from this approximation, for the case of maximal mixing ($\theta_\text{mix} \sim \pi/4$), it is not enough for the mass matrix element $m_{\tilde{\tau}_{12}}$ to be large, but  it is also required  that $m_{\tilde{\tau}_{11}}\sim m_{\tilde{\tau}_{22}}$. In this case  of maximal $\theta_\text{mix}$ there is a cancellation between the couplings of $H$ to  $\tilde{\tau}^{*}_{1}\tilde{\tau}_2$ and to  $\tilde{\tau}_{1}\tilde{\tau}^{*}_{2}$, while the couplings to $\tilde{\tau}_{1}\tilde{\tau}^{*}_{1}$ and $\tilde{\tau}_{2}\tilde{\tau}^{*}_{2}$ are maximal. Since in this situation the decays into pairs of heavier staus  $\tilde{\tau}_2$ are  usually not kinematically available, the decay of $H$ is dominated by the decays into  $\tilde{\tau}_{1}\tilde{\tau}^{*}_{1}$.  The other situation arises when the mixing angle is small, $\theta_\text{mix}\sim 0$, but $ m_{\tilde{\tau}_{12}}$ is still somewhat large, mainly from the large $A_{\tau}$ term~\cite{Djouadi:1996pj}. In this case  the chiral couplings are maximized, such that the left-right part of the coupling of $H$ to  $\tilde{\tau}^{*}_{1}\tilde{\tau}_2$ and the right-left part of the coupling of $H$ to $\tilde{\tau}_{1}\tilde{\tau}^{*}_2$ are maximal. This latter pattern of decay also shows up for the supersymmetric decays of the CP-odd Higgs $A$ to staus due to CP conservation. In this paper we concentrate on the first case and leave the second case for future work.

It is interesting to compare to what is obtained for EW stau-pair production via gauge bosons at 13 TeV, see Figure 1 of~\cite{ATLAS:2016PUB}. Given the final states under consideration, we estimate that the EW cross section is at most of order $10^{-3}$ pb, which is an order of magnitude smaller to the cross sections via heavy Higgs decays. Furthermore, because of the vectorial nature of the gauge bosons couplings, we expect them to be further suppressed due to mixing than the chiral couplings we are considering. Thus we can safely neglect any interference effect and concentrate on production of stau pairs via heavy Higgs decays.

To sum up, within the parameter space region of our interest ($m_H$ $\in$ [800 GeV, 1200 GeV] and $\tan\beta$ $\in$ [25, 50]), the fact of having large stau mixing allow us to obtain BR($H \to \tilde\tau_1 \tilde\tau_1^\ast$) $\sim 0.1-0.2$, which reduces the constraints imposed by the searches for the heavy neutral MSSM Higgs bosons in the di-tau channel. In these scenarios with large $\tan\beta$, the dominant $H$-production mode is by far the $b$-quark annihilation and the stau-pair production cross sections, via $H$ decays, are between one and two orders of magnitude larger than the usual EW production. This situation provides the possibility of stau-pair production and decay through the process $b \bar b$ $\to$ $H$ $\to$ $\tilde\tau_1 \tilde\tau_1^\ast$ $\to$ $\tau\tilde\chi_1^0\tau\tilde\chi_1^0$, with sizable cross sections. Therefore, the final state originated from this decay chain consists of a $\tau$-lepton pair and large $E_T^{\text{miss}}$, with low jet activity.


\section{Collider analysis}
\label{analysis}

In this section we describe an alternative search strategy for the production of a pair of staus through the heavy CP-even Higgs boson $H$ decay, with both scalar fermions decaying subsequently into a tau lepton and the LSP. Hence, the final state involves two opposite-sign tau leptons and large missing transverse energy $E_T^{\text{miss}}$ arising from the pair of LSPs, which escape without being detected. We will detail first the general features of the signal, the kinematic cuts to reduce the main SM backgrounds and the procedure used to optimize its potential discovery at the LHC. After that, in order to broaden the scope of our results, we test the sensitivity of our signal region (SR) by varying the parameters $m_H$, $\tan\beta$, $A_{\tau}$ y $m_{\tilde{\tau}}$, within the context of the large stau-mixing MSSM scenario, providing a broader picture  for our search strategy.

\subsection{Search strategy for stau-pair production via heavy scalar decay}
\label{strategy}

In order to develop our search strategy, we work with a benchmark that possesses the following relevant SUSY parameters: $m_A$ = 947.5 GeV, $\tan\beta =$ 33.8, $M_1$ = 100 GeV, $M_2$ = $M_3$ = 2200 GeV, $\mu$ = $-327.2$ GeV, $A_\tau$ = $-859.4$ GeV, $m_{{\tilde L}_3}$ = 412.9 GeV, and $m_{{\tilde E}_3}$ = 393.8 GeV. This SUSY parameters give rise to the following spectrum (computed with {\tt Spheno 3.3.8}~\cite{Porod:2003um,Porod:2011nf}) for the variables of collider interest: $m_H =$ 947.6 GeV, $m_{\tilde\tau_1} =$ 367.5 GeV, $m_{\tilde\tau_2} =$ 408.4 GeV, and $m_{\tilde\chi_1^0} =$ 99 GeV\footnote{This benchmark point is slightly in tension with the latest $H/A\to\tau\bar{\tau}$ searches~\cite{Aaboud:2017sjh}. Nonetheless, we use it as a reference point to guide us in the construction of the search strategy for staus via heavy Higgs decays. An appropriate allowed point would require a new scan of the MSSM parameter space as the one performed in~\cite{Medina:2017bke}, which is beyond the scope of our work. We expect however that imposing the latests ditau constraints would move $m_H$ to slightly larger values, leaving the rest of the parameters with little modifications.}. For a center-of-mass energy of 14 TeV, the $H$-production cross section via gluon fusion is $\sigma_{ggH} =$ 3.2 fb and via $b$-quark annihilation is $\sigma_{bbH} =$ 194.2 fb, computed both at NNLO with {\tt SusHi}~\cite{Harlander:2012pb,Harlander:2016hcx}, which uses results of~\cite{Harlander:2002wh,Harlander:2003ai,Aglietti:2004nj,Bonciani:2010ms,Degrassi:2010eu,Degrassi:2011vq,Degrassi:2012vt,Harlander:2005rq,Harlander:2003bb,Harlander:2004tp,Harlander:2003kf,Chetyrkin:2000yt}. Hence, it is a good approximation to neglect the former and take into account only the latter. On the other hand, we have BR($H \to \tilde\tau_1 \tilde\tau_1^\ast$) $=$ 0.17 and BR($\tilde\tau_1 \to \tau \tilde\chi_1^0$) $=$ 0.98. Therefore, the total cross section for the complete process $pp$ $\to$ $H$ $\to$ $\tilde\tau_1 \tilde\tau_1^\ast$ $\to$ $\tau\tilde\chi_1^0\tau\tilde\chi_1^0$ is 31.7 fb.

The main backgrounds are listed in Table~\ref{tab_bkg}, where we include the cross sections, estimated by using {\tt MadGraph\_aMC@NLO 2.6}~\cite{Alwall:2014hca} and the number of generated events. 
Although all the events corresponding to the background processes have been generated at leading order, the cross sections for $t\bar{t}$, $WW$ and $ZZ$ have been rescaled with K-factors of 1.5, 1.4, and 1.3, respectively, extracted from~\cite{Alwall:2014hca}. In addition, the cross sections for the $W$+jets and $Z$+jets backgrounds have been estimated by considering up to two light jets. It is important to note that for the $t \bar t$, $W$+jets, and $WW$ backgrounds we have included only the decay of the $W$ boson into $\tau\nu_{\tau}$, while in the case of the $ZZ$ and $Z$+jets backgrounds, we have considered the decays $ZZ\to \tau^+\tau^-\nu\bar{\nu}$  and $Z\to \tau^+\tau^-$, respectively. The multijet QCD background is not taken into account for the signal optimization since it is largely suppressed by the cuts applied on the variables which involve missing transverse energy that we introduce below. Both the signal and the different backgrounds have been generated with {\tt MadGraph\_aMC@NLO 2.6}~\cite{Alwall:2014hca} and showered with {\tt PYTHIA 8}~\cite{Sjostrand:2014zea}, while the detector response has been simulated with {\tt Delphes 3}~\cite{deFavereau:2013fsa}. The implementation of the different cuts of the search strategy that we present below have been also carried out with {\tt MadAnalysis5}~\cite{Conte:2014zja} in the expert mode.\par

\begin{figure}[H]
\begin{center}
\centering
\includegraphics[scale=0.55]{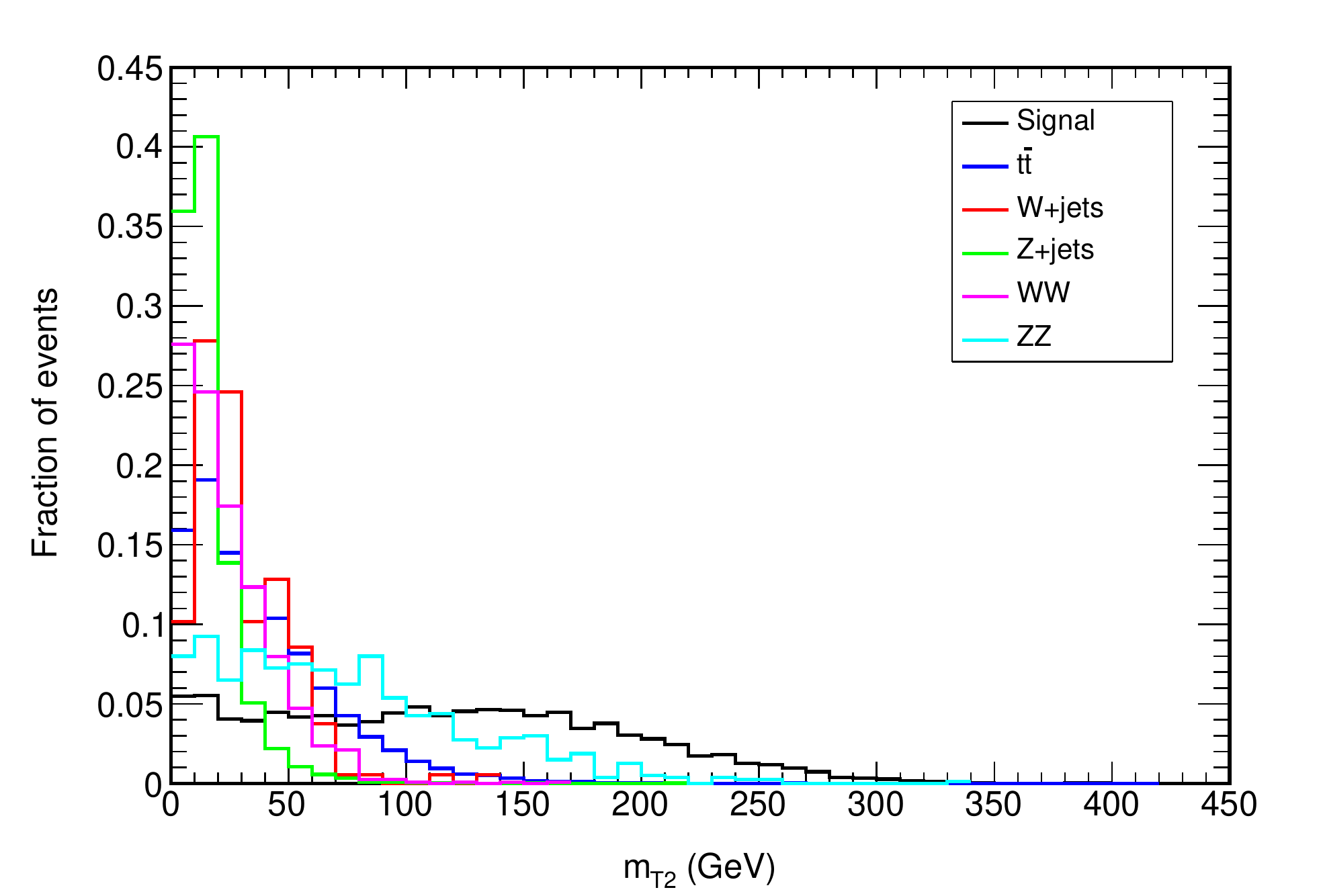}
\caption{Distribution of the $m_{T2}$ variable for the signal and the main backgrounds after requiring exactly two opposite-sign tau leptons in the event that pass the selection cuts.}
\label{plotmt2}
\end{center}
\end{figure}

\renewcommand{\arraystretch}{1.2}
\begin{table}[t]
\vspace*{4mm}
\begin{center}
\begin{tabular}{c|c|c}
\hline \hline
{\bf Background} & {\bf Cross section (fb)} & {\bf Simulated events} \\ \hline\hline 
$t\bar{t}$ & 10125 & $10^{6}$ \\
$W\text{+jets}$ & $6.257\times 10^6$ & $10^{6}$ \\ 
$Z\text{+jets}$ & $4.254\times 10^6$ & $10^{6}$ \\ 
$WW$ & 1188.6  & $1.5\times 10^{5}$ \\ 
$ZZ$ & 183.3 & $10^{5}$ \\  \hline \hline
\end{tabular}
\caption{Main backgrounds along with the corresponding cross sections and the number of simulated events used for this work.}
\label{tab_bkg}
\end{center}
\end{table}
\vspace*{2mm}
We first require that both the signal and background events exhibit exactly two opposite-sign tau leptons, and satisfy the following set of selection cuts:
\begin{equation}
\label{selection}
p_T^{\tau_1} > 50 \, \text{GeV} \,\,,\,\, p_T^{\tau_2} > 40 \, \text{GeV} \,\,, \,\,|\eta^{\tau}| < 2.47 \,,
\end{equation}
where $\tau_1$ ($\tau_2$) denotes the leading (sub-leading) tau lepton, and $\eta^{\tau}$ is the tau pseudo-rapidity. For the final state topology that we are considering here, a powerful discriminating variable commonly used is the $m_{T2}$, which depends on the momenta of two visible particles and the missing transverse energy present in the event. This variable is defined as follows
\begin{equation}
\label{mt2}
m_{T2}=\underset{\pvsslash_{\,\,1}+\pvsslash_{\,\,2}=\vec{p}^{\text{ miss}}_T}{\text{min}}\left\{ \text{max}\left[m_T(\vec{p}^{\,a}_T,\pvslash_{\,1}),m_T(\vec{p}^{\,b}_T,\pvslash_{\,2})\right]\right\} \,,
\end{equation}
where $a$ and $b$ are the two visible states from the parents decays, $\pvslash_{\,1}$ and $\pvslash_{\,2}$ are the corresponding missing momenta, and $\vec{p}^{\text{ miss}}_T$ is the total missing transverse momentum. Finally, the transverse mass $m_T$ is defined as
\begin{equation}
\label{mt}
m_T(\vec{p}^{\,x}_T,\vec{p}^{\text{ inv}}_T)=\sqrt{m^2_x+2(\sqrt{m^2_x+|\vec{p}^{\,x}_T|^2}E^{\text{inv}}_T-\vec{p}^{\,x}_T\cdot\vec{p}^{\text{ inv}}_T)} \,,
\end{equation}
where $x$ denotes the detected particle and $\vec{p}^{\,x}_T$ is its transverse momentum. Since the main feature of the $m_{T2}$ variable is that its distribution has an endpoint around the mass of the parent decaying particle, it is expected that this variable will be very efficient to separate the signal from the $t\bar{t}$ and $WW$ backgrounds. In Figure~\ref{plotmt2} we show the distribution of the $m_{T2}$ variable both for the signal and the various backgrounds, listed in Table~\ref{tab_bkg}, after applying the selection cuts introduced above. We see that this variable is a powerful discriminator not only for the $t\bar{t}$ and $WW$ background processes but also for the $W$+jets, $Z$+jets, and, to a lesser extent, the $ZZ$ background. Taking advantage of the fact that the $m_{T2}$ distributions of $W$+jets, $Z$+jets and $t\bar{t}$ backgrounds exhibit a similar behavior, the acceptances of the first two backgrounds resulting from a given $m_{T2}$ cut have been conservatively assumed to be the one obtained for the $t\bar{t}$ process. 
Besides $m_{T2}$, we have also used other variables with good discrimination power: the angular separation between the two tau leptons in the event, $\Delta R(\tau_1,\tau_2)$, the transverse masses $m_{T\tau_1}$ and $m_{T\tau_2}$ (see Eq.~(\ref{mt})), and the invariant mass of the pair of tau leptons, $m_{\tau\tau}$. In addition, we have imposed requirements in the number of $b$-jets ($N_{b}$) and light jets ($N_{j}$) in the event. The resulting signal region is presented in Table~\ref{tab_SR}. In order to obtain the significance achieved in the proposed signal region, we have assumed a systematic uncertainty of 30\% on the estimated sum of all backgrounds, in accordance to the uncertainties considered in similar searches~\cite{ATLAS:2016PUB,Aad:2014yka}. Including the potential systematic uncertainties, the signal significance can be computed as~\cite{Cowan:2012}
\begin{equation}
\label{Sigsys}
{\cal S} = \sqrt{2 \left((B+S) \log \left(\frac{(S+B)(B+\sigma_{B}^{2})}{B^{2}+(S+B)\sigma_{B}^{2}}\right)-\frac{B^{2}}{\sigma_{B}^{2}}\log \left(1+\frac{\sigma_{B}^{2}S}{B(B+\sigma_{B}^{2})} \right) \right)} \,,
\end{equation}
where $S$ is the number of signal events, $B$ the total number of background events and $\sigma_{B}=(\Delta B) B$, with $\Delta B$ being the relative systematic uncertainty. The results obtained for a total integrated luminosity of 100 fb$^{-1}$ are displayed in Table~\ref{tab_result}. Some comments about our estimation of the number of events of the $W$+jets and $Z$+jets backgrounds are in order:
\renewcommand{\arraystretch}{1.6}
\begin{table}[h]
\begin{center}
\begin{tabular}{c}
\hline \hline
{\bf SR definition} \\ \hline\hline 
2 OS taus and selection cuts \\
$N_{b}=0\,$ \& $\,N_{j}<2$ \\ 
 $\Delta R(\tau_1,\tau_2)<3.5$\\ 
 $m_{T\tau_1},m_{T\tau_2}>120$ GeV\\ 
 $m_{\tau\tau}>100$ GeV \\
 $m_{T2}>180$ GeV \\  \hline \hline
\end{tabular}
\caption{Summary of the cuts involved in the proposed signal region. The selection cuts were defined in Eq.~(\ref{selection}).}
\label{tab_SR}
\end{center}
\end{table} 
\begin{itemize}
\item As can be seen from Tables~\ref{tab_bkg} and \ref{tab_result}, the maximum number of generated events ($10^6$) is well below the number of expected $W$+jets events at 100 fb$^{-1}$ ($>10^8$). Since the generation of such a huge number of events is beyond the scope of our computational resources, we have instead estimated first the acceptance corresponding to all the cuts in Table~\ref{tab_SR} except for the $m_{T2}$ cut, and then used the acceptance of the $t\bar{t}$ background for this last cut. None of the $10^6$ generated events survive the cuts that define the SR (except the cut in $m_{T2}$), therefore we consider that an upper bound on the acceptance at this level is ${\cal O}(10^{-7})$. On the other hand, the acceptance of the $m_{T2}$ cut for the $t\bar{t}$ background is $\sim 0.01$, which leads to the following estimation of the $W$+jets acceptance in the SR: $10^{-7}\times 0.01 = 10^{-9}$.   
\item In the case of $Z$+jets, with $Z\to \tau^+\tau^-$, we obtain that only one out of $10^6$ events remains after applying all the cuts except $m_{T2}$. Thus, the acceptance at this level is $10^{-6}$, which gives, combined with the $m_{T2}$ acceptance corresponding to $t\bar{t}$, an estimation of $10^{-8}$.
\item Finally, for $Z$+jets, but with the $Z$ decaying into neutrinos, the estimation of the acceptance is entirely similar to the case of $W$+jets explained in the first bullet.
\end{itemize}
\par
Finally notice from Table~\ref{tab_result} that even when a 30\% of systematic uncertainties are included, we obtain a $6.62\sigma$ significance of discovery potential with a total integrated luminosity of only 100 fb$^{-1}$. Moreover, on the naive assumption that the background rates scale as the cross section of the signal~\cite{Djouadi:2015jea}, we obtain that the statistical significance for the proposed SR at 300 fb$^{-1}$ is $8.50\sigma$.
\renewcommand{\arraystretch}{1.6}
\begin{table}[h]
\begin{center}
\begin{tabular}{c|c|c}
\hline \hline
&{\bf No cut} & {\bf SR} \\ \hline\hline 
Signal & 3171 & 28.78\\ \hline
$t\bar{t}$ & 1012500 & 2.03\\ 
$W$+jets & $6.257\times 10^8$ & 0.65\\ 
$Z$+jets & $4.254\times 10^8$ & 1.01\\ 
$WW$ & 118860 & 0\\
 $ZZ$ & 18330 & 0.37\\  \hline 
 $\mathcal{S}$ & $1.0\times 10^{-5}$ & 6.62 \\ \hline\hline
\end{tabular}
\caption{Numbers of signal and background events at the 14-TeV LHC with a total integrated luminosity of 100 fb$^{-1}$.}
\label{tab_result}
\end{center}
\end{table} 

\subsection{General interpretation within the large stau-mixing MSSM scenario}
\label{indepe}

\begin{figure}[t]
\begin{center}
\begin{tabular}{cc}
\centering
\hspace*{-3mm}
\includegraphics[scale=0.55]{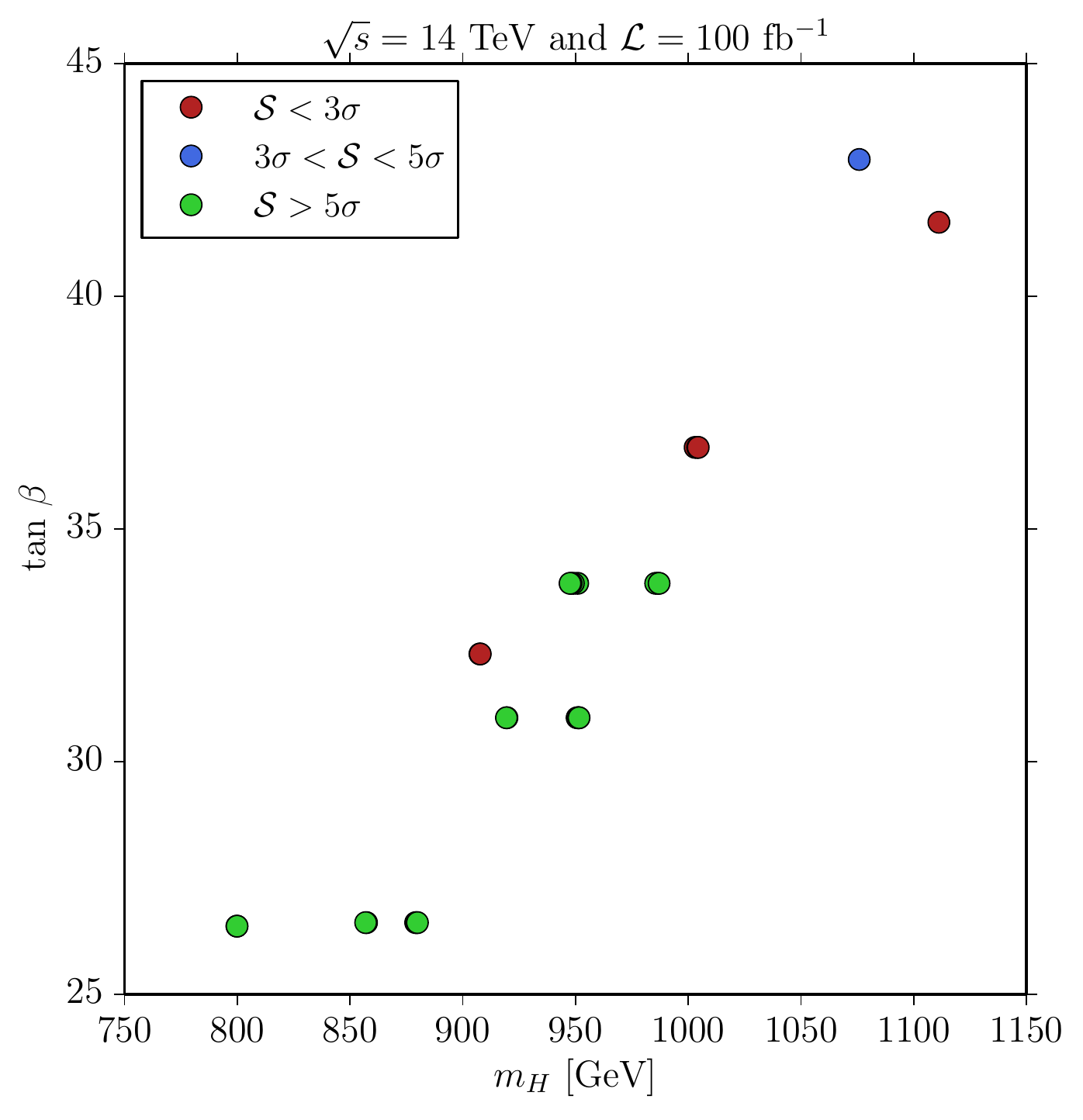} &
\includegraphics[scale=0.55]{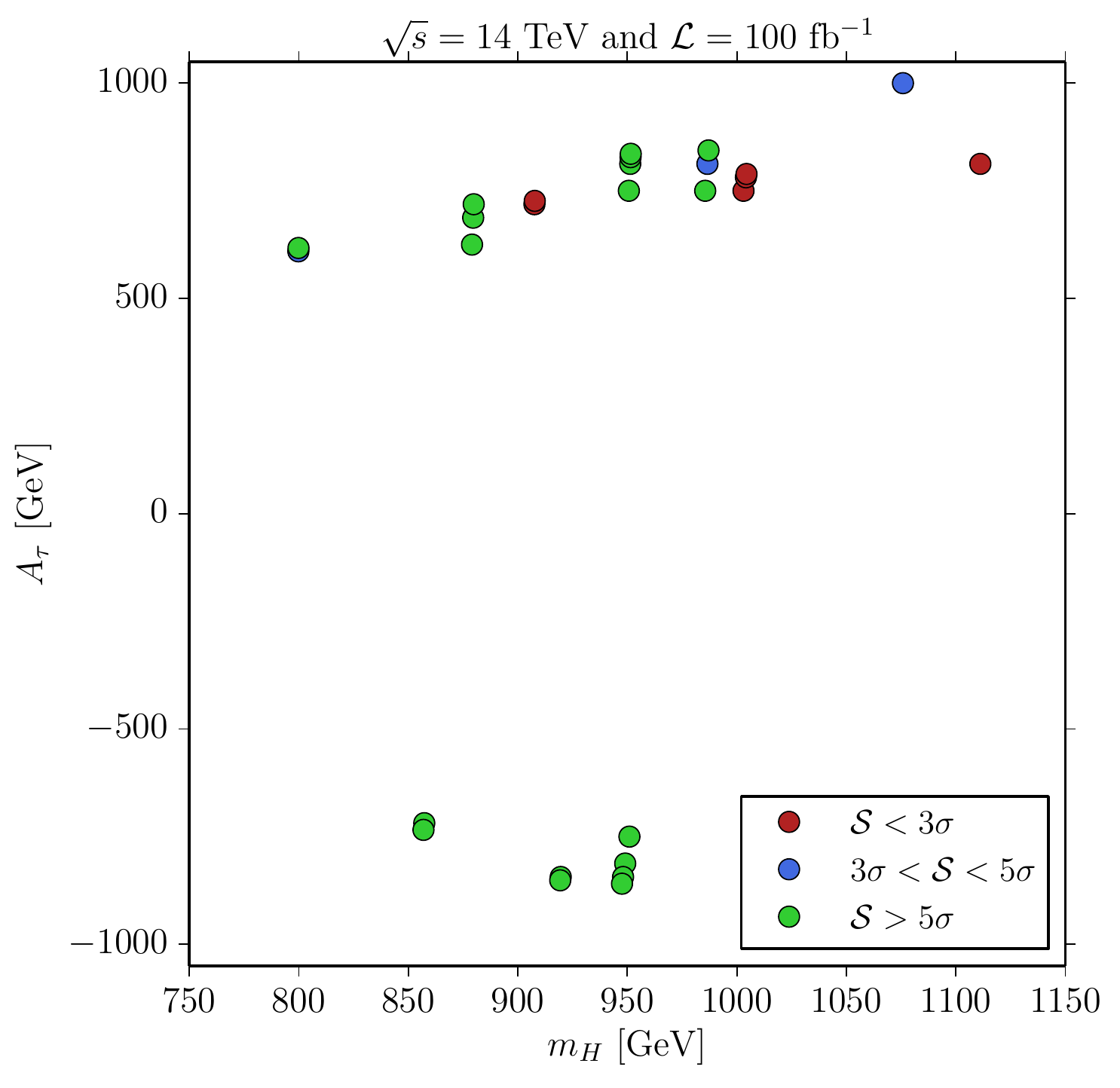}
\end{tabular}
\caption{Signal significance in the planes [$m_H$, $\tan\beta$] (left panel) and [$m_H$, $A_\tau$] (right panel) for various benchmarks within the large stau-mixing MSSM scenario. The displayed values were obtained for a center-of-mass energy of 14 TeV and a total integrated luminosity of 100 fb$^{-1}$. Red circles correspond to significances below the evidence level (${\cal S} < 3\sigma$), blue circles to significances between the evidence level and the discovery one ($3\sigma < {\cal S} < 5\sigma$), and green circles to significances larger than the discovery level (${\cal S} > 5\sigma$).}
\label{points-mHtanb-mHAtau}
\end{center}
\end{figure}

In this section we select 27 benchmarks belonging to the large stau-mixing MSSM scenario, all corresponding to the class of orange points mentioned in Section~\ref{theory} and characterized by different values for $m_H$, $\tan\beta$, $A_\tau$, and $m_{\tilde \tau_1}$. First of all, we have applied to all of them the same search strategy defined in Section~\ref{strategy} and studied the signal significance at the LHC in the planes [$m_H$, $\tan\beta$] and [$m_H$, $A_\tau$], depicted in Figure~\ref{points-mHtanb-mHAtau} for a luminosity $\mathcal{L}=100$ fb$^{-1}$. From both plots of this figure, it is clear that our alternative search strategy for stau pairs at the LHC is very efficient in a broad sense within the context of the large stau-mixing MSSM scenario we work with, despite the fact that this search strategy has been optimized  only for one of the 27 benchmarks considered here. It is important to remember that all the points displayed in the [$m_H$, $\tan\beta$] plane (left panel of Figure~\ref{points-mHtanb-mHAtau}) lie within the region covered by the orange points of Figure 6 of~\cite{Medina:2017bke}.  From this plot we see that the largest significances are obtained for values of $\tan\beta$ below 35, which arises from the fact that the lower $\tan\beta$ is, the lower the branching ratios of $H$ into $b$-quark and $\tau$-lepton pairs are, and then the largest BR($H \to \tilde\tau_1 \tilde\tau_1^\ast$) can be reached. In contrast, for $\tan\beta>35$, the signal significances drop below the discovery level, although it is still possible to reach significances at the evidence level, depending on the values of $A_\tau$ and $m_{\tilde \tau_1}$.
On the right panel of Figure~\ref{points-mHtanb-mHAtau}, the results of our search strategy are shown in the [$m_H$, $A_\tau$] plane, in which we can see that for most of the benchmarks significances above 5$\sigma$ are obtained. Remarkably, our search strategy provides significances at the discovery level for all the benchmarks with negative values of the trilinear $A_\tau$, due to the fact that a negative $A_\tau$ implies a larger stau mixing and in consequence, larger values of BR($H \to \tilde\tau_1 \tilde\tau_1^\ast$). For positive values of $A_\tau$, it is still possible to obtain $\mathcal{S}>5\sigma$ providing $m_H$ is below 1000 GeV, since otherwise the suppression in the $H$ production cross section leads to a decrease in the significance, which can fall even below the evidence level if $m_H$ is large enough.  

\begin{figure}[t]
\begin{center}
\begin{tabular}{cc}
\centering
\hspace*{-3mm}
\includegraphics[scale=0.55]{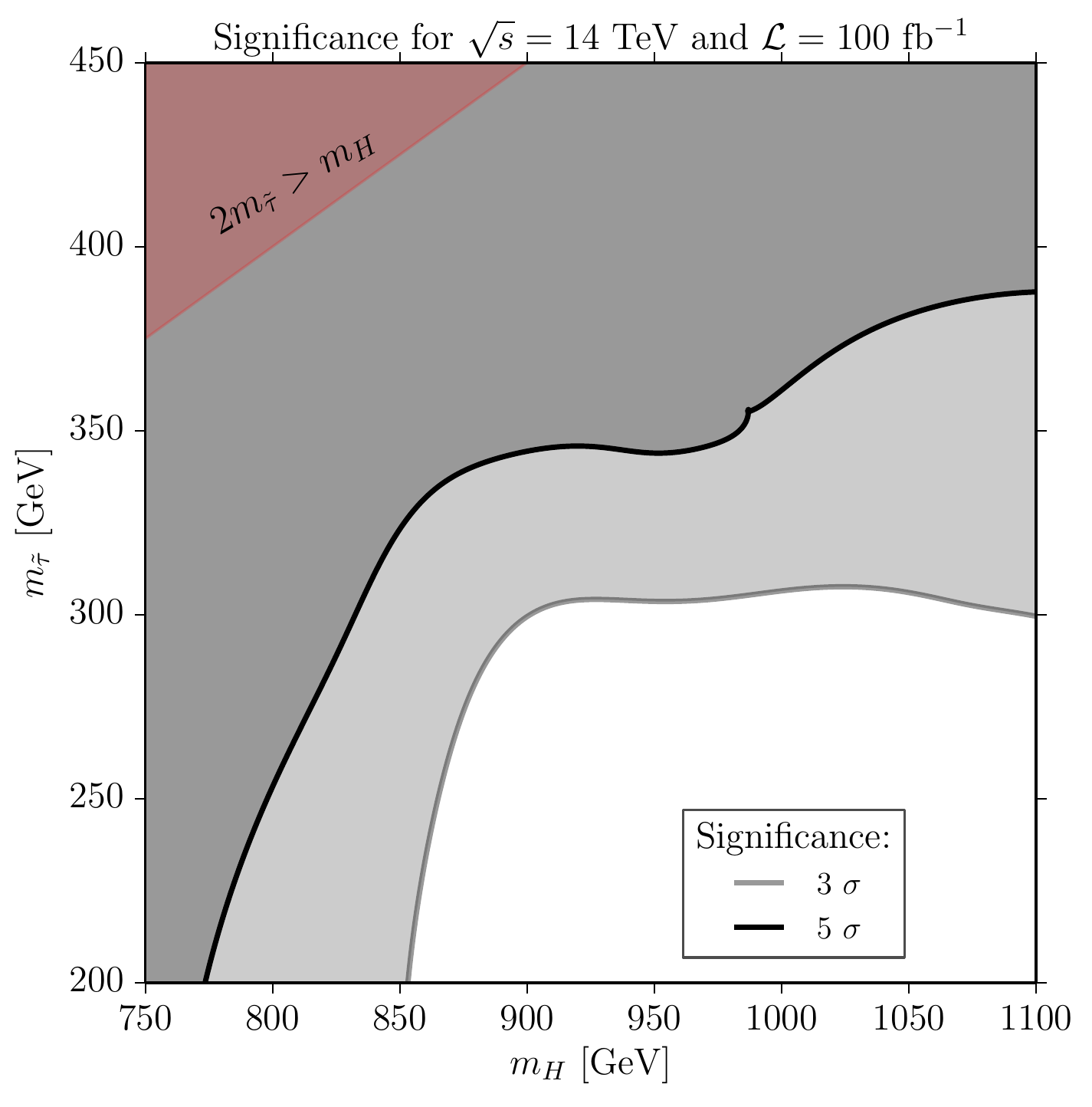} &
\includegraphics[scale=0.55]{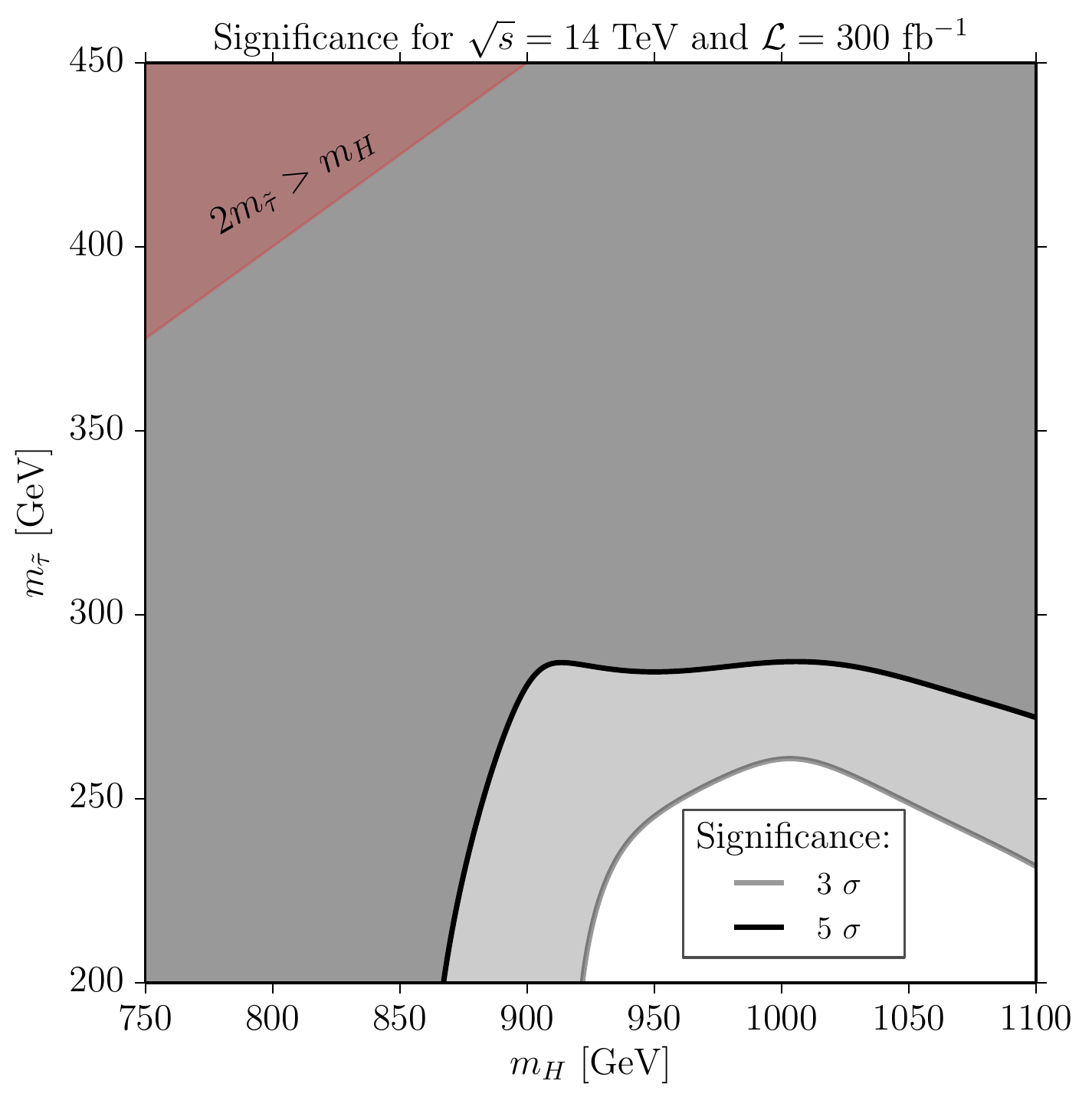}
\end{tabular}
\caption{Signal significance in the [$m_H$, $m_{\tilde \tau_1}$] plane, within the large stau-mixing MSSM scenario, for a center-of-mass energy of $\sqrt{s} =$ 14 TeV and total integrated luminosities of 100 fb$^{-1}$ (left panel) and 300 fb$^{-1}$ (right panel). The light gray area corresponds to significances at the evidence level (3 standard deviations) and the dark gray area to significances at the discovery level (5 standard deviations). Finally, the white area represents signal significances below 3$\sigma$ and the shaded red area is forbidden because the decay mode $H \to \tilde\tau_1 \tilde\tau_1^\ast$ is kinematically closed.}
\label{contour-mstautanb}
\end{center}
\end{figure}

It is also interesting to interpret our results by performing an extrapolation  in the [$m_H$, $m_{\tilde \tau_1}$] plane from the 27 benchmark points, as displayed in the contours of Figure~\ref{contour-mstautanb}. For a total integrated luminosity of 100 fb$^{-1}$ (left panel of Figure~\ref{contour-mstautanb}), there is a big portion of the parameter space ($m_H >$ 850 GeV and $m_{\tilde \tau_1} <$ 300 GeV) in which our proposed search strategy would not be sensitive to this signal (white area).  This situation occurs for two reasons: on the one hand, the signal cross section is considerably reduced  for large values of $m_H$, and on the other one, small stau masses produce a final state with less energetic tau leptons and lower $E_T^\text{miss}$, which in turn reduce the discrimination power of crucial kinematic variables as $m_{T2}$ or $m_T$.
Conversely, if $m_H$ is reduced or $m_{\tilde \tau_1}$ increased, the significances enter first into the evidence level (light gray area) and later on into the discovery level (dark gray area). Interestingly, for values larger than 350 GeV, we obtain significances above 5$\sigma$ for practically any value of the heavy scalar mass, except for very large values of $m_H$, in which case we would need $m_{\tilde \tau_1} >$ 400 GeV.
If one considers a total integrated luminosity of 300 fb$^{-1}$ (right panel of Figure~\ref{contour-mstautanb}), the white area in which the proposed search strategy is not sensitive is substantially reduced ($m_H >$ 925 GeV and $m_{\tilde \tau_1} <$ 250 GeV). In fact, for this luminosity our search strategy is sensitive to the signal in most of the parameter space, with significances at the discovery level for $m_H <$ 850 GeV or $m_{\tilde \tau_1} >$ 300 GeV.

\section{Conclusions}
\label{conclusions}
In this work we proposed, within the context of the MSSM, a novel search strategy for staus at the LHC  based on the resonant s-channel heavy CP-even Higgs boson production via $b$-quark annihilation, which can be significantly large for $\tan\beta\gg 1$.
In certain regions of the parameter space analyzed which safely satisfy all collider bounds, there can be a sizable branching ratio of $H$ into staus, leading to a stau-pair production cross section that is between one or two orders of magnitude larger than the usual EW production for these sfermions. We first presented the MSSM scenarios with large stau mixing, in which it is possible to obtain large values of BR($H \to \tilde\tau_1 \tilde\tau_1^\ast$) $\sim 0.1-0.2$, that allow at the same time to safely avoid the strong constraints in the [$m_A$, $\tan\beta$] plane from the current ATLAS and CMS searches for $H$ and $A$ in the di-tau channel. We then detailed our search strategy for this class of stau-pair production, focusing in the case when both staus decay into a tau lepton and the LSP, the lightest neutralino.  In this way, the experimental signature of this SUSY signal is made up of a tau-lepton pair and large $E_T^{\text{miss}}$ arising from the two LSPs. In order to increase the signal-over-background ratio, we performed a set of cuts in different kinematic variables generally used in the literature, among which the $m_{T2}$ variable is the most discriminant for the particular final state topology considered here. We obtained significances of discovery potential (larger than 5$\sigma$) for a center-of-mass energy of $\sqrt{s} =$ 14 TeV with only a total integrated luminosity of 100 fb$^{-1}$, even under the conservative assumption of a 30\% of systematic uncertainties in the total background.

Finally, we  generalized the results by applying the proposed search strategy to several other benchmarks that also belong to the large stau-mixing scenario and are represented  by the parameters $m_H$, $m_{\tilde \tau_1}$, $\tan\beta$, and $A_{\tau}$. In this case the results are also promising, with the significance being above the discovery level for most of the tested benchmarks for luminosities attainable in the near future. The proposed search strategy appears to be highly sensitive in most of the parameter space, except when the heavy scalar mass is very large, typically above 925 GeV, and/or the stau mass is small, approximately below 250 GeV. The sensitivity also worsens for values of $\tan\beta$ above $\sim$ 35 due to the increment in the values of the branching ratios of $H$ into $b$-quark and $\tau$-lepton pairs that takes place in this case. Nonetheless, it is still possible to achieve significances at the evidence level for certain values of the trilinear $A_{\tau}$.

We have shown in this work that the search for stau-pair production at the LHC coming from the heavy CP-even scalar $H$ decay is very promising in regions of the MSSM still allowed by experiments, with a much better prospect than the usual stau-pair searches based on the EW production.

\section*{Acknowledgments}
We thank Carlos Wagner and Michael A. Schmidt for useful discussions and insight. E.A. warmly thanks IFT of Madrid for its hospitality hosting him during the completion of this work. V.M.L. thanks Xabier Marcano and Claudia Garc\'ia Garc\'ia for fruitful conversations. This work has been partially supported by CONICET and ANPCyT project no. PICT 2016-0164
 (E.A., A.M., and N.M.). V.M.L. acknowledges support of the
SPLE ERC project and the BMBF under project 05H15PDCAA. 

\bibliographystyle{unsrt}

\end{document}